\documentclass[aps,prd,preprint,groupaddress,amssymb,amsmath,nofootinbib]{revtex4}

\usepackage{color}
\usepackage{graphicx}
\usepackage{float}
\usepackage[usenames,dvipsnames]{xcolor}

\begin{document}

\preprint{}

\title{Top-bottom-tau Yukawa coupling unification in the MSSM plus one vectorlike family and fermion masses as IR fixed points}
%\title{Top-bottom-tau Yukawa coupling unification in the MSSM+1VF and their IR fixed points}
%\title{IR fixed point nature of top-bottom-tau Yukawa coupling unification in the MSSM+1VF}
%\title{Top, bottom and tau masses as IR fixed points from unified Yukawa coupling in the MSSM+1VF}

%\title{Top, bottom and tau masses from IR fixed points  in the MSSM+1VF}
%\title{Third generation fermion masses from IR fixed points  in the MSSM+1VF}

\author{Radovan Derm\' \i\v sek}

\email[]{dermisek@indiana.edu}

\author{Navin McGinnis}

\email[]{nmmcginn@indiana.edu}

\affiliation{Physics Department, Indiana University, Bloomington, IN 47405, USA}

%\homepage[]{Your web page}
        %\thanks{}
%\altaffiliation {}

%\date{\today}
\date{October 19, 2018}

\begin{abstract}

In the MSSM extended by a complete vectorlike family, precise top, bottom and tau Yukawa coupling unification can be achieved assuming SUSY threshold corrections which are typical for comparable superpartner masses. Furthermore, the unification is possible with a large unified coupling, implying that all three fermion masses can be simultaneously close to their IR fixed points. Assuming unified Yukawa couplings of order one or larger, the preferred common scale of new physics (superpartners and vectorlike matter) is in the 3 TeV to 30 TeV range, with larger couplings favoring smaller scales. Splitting superpartner masses from masses of vectorlike fields, the preferred scales extend in both directions. The multi-TeV scale for superpartners  is compatible with and  independently suggested by  the Higgs boson mass.

\end{abstract}

% insert suggested PACS numbers in braces on next line
\pacs{}
% insert suggested keywords - APS authors don't need to do this
\keywords{}

%\maketitle must follow title, authors, abstract, \pacs, and \keywords
\maketitle

% body of paper here - Use proper section commands
% References should be done using the \cite, \ref, and \label commands
%\section{ \label{sec:}}
% Put \label in argument of \section for cross-referencing
%\section{\label{}}

%\subsection{}

%\subsubsection{}

%\subsection{Introduction}

%\subsubsection{Experimental results}

\section{Introduction}
\label{sec:intro}

Values of some of the free parameters in the standard model (SM) can be understood if they are related by additional symmetries to other parameters. Gauge coupling unification in the minimal supersymmetric extension of the standard model (MSSM) is a well known example that points to a larger symmetry of a grand unified theory (GUT) at the scale where gauge couplings meet. Similarly, embedding the particle content of the SM into GUT multiplets offers a possibility to understand Yukawa couplings and thus fermion masses from a unified Yukawa coupling at the same scale. There are indications that at least the masses of the third generation fermions (top quark, bottom quark and tau lepton) can be understood in this way as motivated by SO(10) symmetry~\cite{Ananthanarayan:1991xp, Ananthanarayan:1992cd, Barger:1992ac, Anderson:1992ba, Hall:1993gn, Hempfling:1993kv, Carena:1994bv,  Murayama:1995fn, Blazek:2001sb, Blazek:2002ta, Baer:2001yy, Auto:2003ys, Tobe:2003bc, Balazs:2003mm, Baer:2009ie, Nath:2009nf, Badziak:2011wm, Baer:2012cp, Joshipura:2012sr, Elor:2012ig, Baer:2012jp, Anandakrishnan:2012tj, Ajaib:2013zha, Karozas:2017hog, Poh:2017xvg, Dutta:2018yos}.

However, the predictive power of Yukawa coupling unification is reduced because  other (so far unknown) parameters also enter the determination of fermion masses. In the MSSM, the crucial parameter is the ratio of vacuum expectation values of the two Higgs doublets, $\tan\beta$, that sets the required top, bottom and tau Yukawa couplings from their measured masses.  Furthermore, there are significant supersymmetric (SUSY) threshold corrections~\cite{Hall:1993gn, Hempfling:1993kv, Carena:1994bv, Blazek:1995nv}  that, in the range of $\tan \beta$ favored by Yukawa coupling unification, can comprise up to about half of the bottom quark mass depending on superpartner masses. Without knowing $\tan\beta$ and at least  basic features of SUSY spectrum, there is no sharp prediction for fermion masses. Nevertheless, we  can instead  require that the third generation of fermion masses  originate from a single Yukawa coupling at the GUT scale and predict $\tan\beta$ and the SUSY spectrum consistent with this assumption. This has been done in a variety of scenarios~\cite{Blazek:2001sb, Blazek:2002ta, Baer:2001yy, Auto:2003ys, Tobe:2003bc, Balazs:2003mm, Baer:2009ie, Nath:2009nf, Badziak:2011wm, Baer:2012cp, Joshipura:2012sr, Elor:2012ig, Baer:2012jp, Anandakrishnan:2012tj, Ajaib:2013zha}  typically pointing to certain  hierarchies or relations among SUSY parameters. 

The predictive power of Yukawa coupling unification can be increased if the required electroweak (EW) scale values of Yukawa couplings  are close to the IR fixed points of Yukawa couplings in a given model.  This makes the actual value of the unified Yukawa coupling unimportant, effectively reducing the number of relevant model parameters to two:  $\tan\beta$ and the SUSY threshold correction to the bottom quark mass.  Although such a possibility does not work in the MSSM,\footnote{In the MSSM the top quark mass can be understood from the IR fixed point value of the top Yukawa coupling~\cite{Bardeen:1993rv, Carena:1993bs, Carena:1994ax}. However, it requires small $\tan \beta$ precluding Yukawa coupling unification. For large $\tan \beta$, the top Yukawa coupling is below the IR fixed point and approaches it very slowly in the RG evolution.} we will see that it works very well in the MSSM extended by a complete vectorlike family (an exact copy of a SM family: $q$, $\bar u$, $\bar d$, $l$, $\bar e$ and corresponding fields with conjugate quantum numbers). 

We show that in the MSSM extended by a complete vectorlike family (MSSM+1VF), precise top, bottom and tau Yukawa coupling unification can be achieved with a large unified coupling, implying that all three fermion masses can be simultaneously close to their IR fixed points. All three Yukawa couplings approach IR fixed points rapidly from a large range of boundary conditions both above and below the IR fixed point values. Furthermore, the unification is possible assuming SUSY threshold corrections which are typical for comparable superpartner masses and thus no hierarchies or specific relations among SUSY parameters are required. Assuming unified Yukawa couplings of order one or larger, the preferred common scale of new physics (superpartners and vectorlike matter) is in the 3 TeV to 30 TeV range, with larger couplings favoring smaller scales. Splitting superpartner masses from masses of vectorlike fields, the preferred scales extend in both directions.  The required scale of new physics is to a large extent  driven by fitting the measure values of gauge couplings~\cite{Dermisek:2017ihj} with fermion masses further constraining the preferred range. However, due to the IR fixed point behavior it is highly non-trivial that  Yukawa couplings point to a similar scale of new physics as gauge couplings. Furthermore, the multi-TeV scale for superpartners  is compatible with and  independently suggested by  the Higgs boson mass.

Enlarging the particle content of the model also results in new parameters that cast a shadow on the predictivity of Yukawa coupling unification. The fields in a vectorlike family can have Yukawa couplings to Higgs doublets, they can mix with SM families (we will not consider this possibility) and they must have vectorlike masses to avoid detection. However we will see that there are only three parameters: the GUT scale, the scale of new physics (superpartner masses and masses of vectorlike matter) and $\tan\beta$ that are important for the EW scale values of standard model Yukawa couplings while others affect the EW scale values very little and are only needed for precisely reproducing the measure values. Furthermore, two of these parameters, the GUT scale and the scale of new physics, are independently constrained by measured values of gauge couplings. 

We assume the common scale of new physics only  for simplicity. The results do not differ much as long as superpartner masses and vectorlike masses are comparable. Nevertheless, after presenting the main results, we will also explore effects resulting from abandoning our simple assumptions. We will consider the scale of superpartners independent from vectorlike masses. Moreover, since the assumption of a common scale for superpartners has an impact on the predicted bottom quark mass, through SUSY threshold corrections, we will  also consider splitting gaugino masses from scalar masses. Furthermore, we will present results in terms of the required SUSY correction to the bottom quark mass that could be used in a variety of scenarios that are not approximated well by our assumptions.

Extensions of the SM or the MSSM with vectorlike   matter were previously explored in a variety of contexts. 
Examples  include studies of their effects on gauge couplings~\cite{Moroi:1993, Babu:1996zv, Kolda:1996ea, Ghilencea:1997yr, AmelinoCamelia:1998tm, BasteroGil:1999dx, Dermisek:2012as, Dermisek:2012ke},~\cite{Dermisek:2017ihj} and on electroweak symmetry breaking and the Higgs boson mass~\cite{Babu:2008ge, Martin:2009bg, Dermisek:2016tzw}. In addition, vectorlike fermions are often introduced on purely phenomenological grounds to explain various anomalies. Examples include discrepancies in precision Z-pole observables~\cite{Choudhury:2001hs, Dermisek:2011xu, Dermisek:2012qx, Batell:2012ca} and the muon g-2 anomaly~\cite{Kannike:2011ng, Dermisek:2013gta} among many others. 
More related to our study, the fast approach of Yukawa couplings to the IR fixed points in asymptotically divergent models was observed in Refs.~\cite{Bando:1997dg},~\cite{Dermisek:2012as, Dermisek:2012ke},~\cite{Dermisek:2017ihj} and the $b-\tau$ Yukawa coupling unification in the MSSM with vectorlike matter was recently discussed in Ref.~\cite{Chigusa:2017drd}.

This paper is organized as follows. In Sec.~\ref{sec:RGE},  we discuss model parameters and assumptions, provide approximate formulas for the RG equations of Yukawa couplings and SUSY threshold corrections and summarize details of the numerical analysis. The  main results and their discussion are contained in Sec.~\ref{sec:results} and we conclude in Sec.~\ref{sec:conclusions}.

\section{Model parameters, RG equations and procedure}
\label{sec:RGE}

We start exploring predictions for top, bottom and tau Yukawa couplings with the following set of model parameters: 
\begin{equation}
M_G, \; M, \; \tan \beta,  
\label{eq:mass_pars}
\end{equation}
representing the GUT scale, the common mass of vectorlike matter and superpartners, $M\equiv M_{V} = M_{SUSY}$, and the ratio of vacuum expectation values of the two Higgs doublets, $\tan \beta = v_u/v_d$; together with 
\begin{equation}
 \alpha_G, \; \epsilon, \; Y_0, \; Y_{V},  
 \label{eq:couplings_pars}
\end{equation}
denoting the unified value of gauge couplings at the GUT scale, the GUT scale threshold correction to gauge couplings and the GUT scale boundary conditions for the common Yukawa coupling of top, bottom and tau,
\begin{equation}
 y_t (M_G) = y_b(M_G) = y_\tau(M_G) \equiv Y_0,
 \label{eq:Y0}
\end{equation}
and the common Yukawa coupling of vectorlike matter. We neglect Yukawa couplings of first two SM generations.

We define the GUT scale as the scale where $\alpha_1$ and $\alpha_2$ differ from $\alpha_3$ by equal amounts and we identify $\alpha_G$ with $\alpha_3$ at this scale:
\begin{equation}
\alpha_G = \alpha_3(M_G), \;\;\;\;\;\;\;\;\;\;\;\;\;   \alpha_1 (M_G) = \alpha_G (1-\epsilon), \;\;\;\;\;\;\;\;\;\;\;\;\; \alpha_2 (M_G) = \alpha_G (1+\epsilon).
\end{equation}
This is different than the common definition of the GUT scale as the scale where $\alpha_1(M_G) = \alpha_2(M_G)$. We prefer the above definition since it  places the GUT scale close to the middle of the interval determined by  scales where two individual couplings meet rather than at the lower edge of this interval as  with the common definition. In addition, it is $\alpha_3$ that plays the most important role in the RG evolution of Yukawa couplings. Nevertheless, this choice does not have a significant effect on the presented results.

The RG evolution of the third generation Yukawa couplings will be affected by possible Yukawa couplings of vectorlike fields. Motivated by the possibility of  embedding the whole generation of SM fields into 16 dimensional representation of SO(10) and assuming no mixing between the third generation and vectorlike matter, there can be two unified Yukawa couplings of vectorlike fields at the GUT scale: $Y$ for fields with the same quantum numbers as the SM fields and $\bar Y$ for fields with conjugate quantum numbers. 
The Yukawa part of the superpotential can be summarized as 
\begin{equation}
 W \; \supset \; Y_0 \, 16_3 10_H 16_3 \;+\; Y \, 16 10_H 16  \;+\; \bar Y \, \bar{16} 10_H \bar{16} ,
 \label{eq:W}
\end{equation}
where the third generation SM fields originate from $16_3$, the two Higgs doublets from $10_H$ and the vectorlike fields from $16$ and $\bar{16}$. Below the GUT scale, the Yukawa couplings of individual fields will evolve according to their corresponding RG equations. Labeling the additional couplings of vectorlike fields, $Y$ and $\bar Y$, with subscripts corresponding to individual fields, the 1-loop RG equations for top, bottom and tau Yukawa couplings are:
\begin{eqnarray}
\frac{dy_t}{dt} &=&  \frac{y_t}{16\pi^2} \left( 3y_t^* y_t + y_b^* y_b +T_{H_u} - \frac{16}{3}g_3^2 - 3g_2^2 - \frac{13}{15}g_1^2 \right), \label{eqn:RGtop} \\
\frac{dy_b}{dt} &=&  \frac{y_b}{16\pi^2} \left( 3y_b^* y_b + y_t^* y_t +T_{H_d} - \frac{16}{3}g_3^2 - 3g_2^2 - \frac{7}{15}g_1^2 \right),\label{eqn:RGbottom} \\
\frac{dy_\tau}{dt} &=&  \frac{y_\tau}{16\pi^2} \left( 3y_{\tau}^* y_\tau + T_{H_d} - 3g_2^2 - \frac{9}{5}g_1^2 \right),
\label{eqn:RGtau}
\end{eqnarray} 
where $ t = \ln Q/Q_0$, with $Q$ being the RG scale, and 
\begin{eqnarray}
T_{H_u} &\equiv& 3y_t^* y_t + 3Y_{U}^*Y_{U} + 3\bar Y_{D}^*\bar Y_{D} + \bar Y_{E}^*\bar Y_{E},\\
T_{H_d} &\equiv&  y_{\tau}^* y_\tau + 3y_b^* y_b  + 3\bar Y_{U}^*\bar Y_{U} + 3Y_{D}^*Y_{D} + Y_{E}^*Y_{E},
\end{eqnarray}
represent the sums of Yukawa couplings squared of  all the fields that couple to the corresponding Higgs doublet.  Note that,  because of conjugate quantum numbers, the fields from $\bar{16}$ couple to Higgs doublets in a flipped way compared to fields in $16$. The conjugate down quark and charged lepton from $\bar{16}$ couple to $H_u$ while conjugate up quark couples to $H_d$. The RG equations for  Yukawa couplings of vectorlike fields can be obtained from those above with obvious replacements.  We assume that all SM singlets (right handed neutrinos) remain at the GUT scale and thus do not contribute in the RG evolution to low energies (assuming these fields to be present to an intermediate scale would not have a qualitative impact on presented results).  Furthermore, for simplicity and also not to favor contributions to the top or bottom Yukawa couplings in the RG evolution,  we assume $Y = \bar Y \equiv Y_V$ at the GUT scale.

We will see that the three parameters in Eq.~(\ref{eq:mass_pars}) are the most important for the EW scale values of standard model Yukawa couplings while those in Eq.~(\ref{eq:couplings_pars}) affect the EW scale values very little and are only needed for precisely reproducing the measured values. Identifying $M_{SUSY}$ with the scale of vectorlike matter, $M_V$, is only done for simplicity and the results do not differ much as long as these two scales are comparable (after presenting the main results, we will explore the effects of splitting $M_{SUSY}$ from $M_V$). Similarly, the assumption of a common scale of vectorlike matter does not have a significant impact on the predicted fermion masses. A split spectrum of vectorlike matter (for example the spectrum obtained from the RG evolution starting with a unified vectorlike mass term at the GUT scale, $M_V 16  \bar{16}$) would result in logarithmic threshold corrections to the third generation Yukawa couplings.\footnote{The impact of the assumption of a common vectorlike mass at the EW scale versus the GUT scale  on gauge couplings in this scenario was studied in Ref.~\cite{Dermisek:2017ihj}. The common mass at the GUT scale leads to an improvement in gauge coupling unification. However the difference is not dramatic.} Unless the splitting is significant these effects are small and can be easily compensated for by small changes in other model parameters. Thus we will not consider these possibilities.

However, the assumption of a common scale for superpartners has a significant impact, especially on the predicted bottom quark mass, through finite SUSY threshold corrections~\cite{Hall:1993gn, Hempfling:1993kv, Carena:1994bv, Blazek:1995nv}.
We match the SM top, bottom and tau Yukawa couplings to those in the MSSM+1VF at the $M_{SUSY}$ scale:
\begin{eqnarray}
y_{t,SM} (M_{SUSY}) &=& y_t (M_{SUSY}) \sin \beta \, (1 + \epsilon_t), \label{eq:yt-corrections} \\
y_{b,SM} (M_{SUSY}) &=& y_b (M_{SUSY}) \cos \beta \, (1 + \epsilon_b),\\
y_{\tau,SM} (M_{SUSY}) &=& y_\tau (M_{SUSY}) \cos \beta \, (1 + \epsilon_\tau) \label{eq:ytau-corrections},
\end{eqnarray}
where $\epsilon_{t,b,\tau}$ are SUSY threshold corrections. Typically dominant contributions are from gluino-stop loops for the top quark, 
\begin{equation}
\epsilon_t \simeq \frac{2\alpha_3}{3\pi} \, M_{\tilde{g}}\mu \, \cot\beta \, I(m^2_{\tilde{t},1},m^2_{\tilde{t},2}, M^2_{\tilde{g}}),
\end{equation}
gluino-sbottom and chargino-stop loops for the bottom quark,
\begin{equation}
\epsilon_b \simeq \frac{2\alpha_3}{3\pi}\, M_{\tilde{g}}\mu \, \tan\beta \, I(m^2_{\tilde{b},1},m^2_{\tilde{b},2}, M^2_{\tilde{g}}) + \frac{y_t^2}{16\pi^2}\, A_t\mu \, \tan\beta \, I(m^2_{\tilde{t},1},m^2_{\tilde{t},2}, \mu^2),
\end{equation}
and bino-stau loops for the tau lepton,
\begin{equation}
\epsilon_\tau \simeq \frac{\alpha_1}{4\pi}\, M_{\tilde{B}}\mu \, \tan\beta \, I(m^2_{\tilde{\tau},1},m^2_{\tilde{\tau},2}, M^2_{\tilde{B}}),
\end{equation}
where subscripts 1 and 2 label  two mass eigenstates of corresponding scalars and 
\begin{equation}
I(a,b,c) \equiv \frac{ab\ln(a/b) + bc\ln(b/c) + ac\ln(c/a)}{(a-b)(b-c)(a-c)}.
\end{equation}

%\begin{eqnarray}
%\epsilon_t &=& \frac{2\alpha_s}{3\pi}M_{\tilde{g}}\mu\cot\beta I(m^2_{\tilde{t},1},m^2_{\tilde{t},2}, M^2_{\tilde{g}}),\\
%\epsilon_b &=& \frac{2\alpha_s}{3\pi}M_{\tilde{g}}\mu\tan\beta I(m^2_{\tilde{b},1},m^2_{\tilde{b},2}, M^2_{\tilde{g}}) + \frac{y_t^2}{16\pi^2}A_t\mu\tan\beta I(m^2_{\tilde{t},1},m^2_{\tilde{t},2}, \mu^2),\\
%\epsilon_\tau &=& \frac{\alpha_1}{4\pi}M_{\tilde{B}}\mu\tan\beta I(m^2_{\tilde{\tau},1},m^2_{\tilde{\tau},2}, M^2_{\tilde{B}}),
%\end{eqnarray}
%with
%\begin{equation}
%I(a,b,c) \equiv \frac{ab\ln(a/b) + bc\ln(b/c) + ac\ln(c/a)}{(a-b)(b-c)(a-c)}.
%\end{equation}

The SUSY threshold corrections for the top Yukawa coupling are small in the large $\tan\beta$ region characteristic for Yukawa coupling unification. The corrections are also small for the tau Yukawa coupling since they are  proportional to $\alpha_1$. However, for the bottom Yukawa coupling, they are of order $1\% \times \tan\beta $ and typically in the 30\%--40\% range assuming comparable superpartner masses. In the limit where all superpartner masses are equal, given by $M_{SUSY}$, the chargino correction is an order of magnitude smaller than the gluino correction for A-terms as large as $M_{SUSY}$.\footnote{For very large A-terms the chargino correction can be comparable to gluino correction or even dominate. The region of the parameter space in the MSSM where gluino and chargino corrections almost cancel leading to successful  Yukawa coupling unification was explored in Refs.~\cite{Blazek:2001sb, Blazek:2002ta, Baer:2001yy, Auto:2003ys}.} In addition, whether the chargino correction adds to or subtracts from the gluino correction depends on the relative sign of the A-term and gluino mass and thus, for simplicity we assume zero A-terms when presenting main results. In the limit of degenerate superpartner masses the loop function also simplifies, $I(M^2,M^2,M^2) = 0.5M^{-2}$. Finally,  electroweak symmetry breaking requires  $\mu^2 \simeq - m_{H_u}^2$ and the typical result from the RG flow over few orders of magnitude in the energy scale is $m_{H_u}^2 \simeq - m_{\tilde t_L}^2 - m_{\tilde t_R}^2$, see for example Ref.~\cite{Dermisek:2016tzw}. Thus, the typical expectation is $\mu \simeq \pm \sqrt{2} M_{SUSY}$ with either sign. 
With these assumptions and simplifications  the approximate formulas for the SUSY threshold corrections are:
\begin{eqnarray}
\epsilon_t &\simeq&\frac{\sqrt{2}\alpha_3}{3\pi}\text{sgn}(\mu)\cot\beta,\\
\epsilon_b &\simeq& \frac{\sqrt{2}\alpha_3}{3\pi}\text{sgn}(\mu)\tan\beta,\\
\epsilon_\tau &\simeq& \frac{\sqrt{2}\alpha_1}{8\pi}\text{sgn}(\mu)\tan\beta,
\end{eqnarray}
from which the typical sizes can be readily obtained. For any specific SUSY breaking scenario the SUSY corrections could be evaluated precisely. However, the above formulas  should be a good approximation in large regions of the parameter space of  scenarios with both high and low mediation scales of SUSY breaking. In addition to the main results assuming a common scale of superpartners we will  explore the impact of splitting gaugino masses from scalar masses. Furthermore, we will also present results in terms of the required SUSY correction to the bottom quark mass that could be used in a variety of scenarios that are not approximated well by our assumptions.

In the numerical study we use 3-loop RG equations for gauge couplings and 2-loop RG equations for the third generation Yukawa couplings and Yukawa couplings of vectorlike fields~\cite{Jones:1975, Machacek:1983tz, Machacek:1983fi, Machacek:1984zw, Martin:1993, Castano:1993ri}, \cite{Kolda:1996ea}. All the particles above the EW scale are integrated out at their corresponding mass scales. The complete set of SUSY threshold corrections to the third generation Yukawa couplings (for which the approximate formulas can be found above) is included at the $M_{SUSY}$ scale~\cite{Hall:1993gn, Hempfling:1993kv, Carena:1994bv, Blazek:1995nv, Pierce:1996zz} with the assumption that $\mu = - \sqrt{2} M_{SUSY}$ (we will see that only the negative sign is consistent with Yukawa coupling unification assuming comparable superpartner masses). When fitting the central values of gauge couplings and fermion masses we use as an input:  $\alpha_{EM}^{-1} (M_Z) = 127.955$, $\sin^2 \theta_W = 0.2312$, $\alpha_{3} (M_Z) = 0.1181$, $m_t = 173.1$ GeV, $m_b(m_b) = 4.18$ GeV and $m_\tau = 1.777$ GeV, where $m_t$ and $m_\tau$ are pole masses~\cite{Tanabashi:2018oca}.

 \section{Results}
 \label{sec:results}
  
\begin{figure}[t]
\centering
\includegraphics[width = 4in]{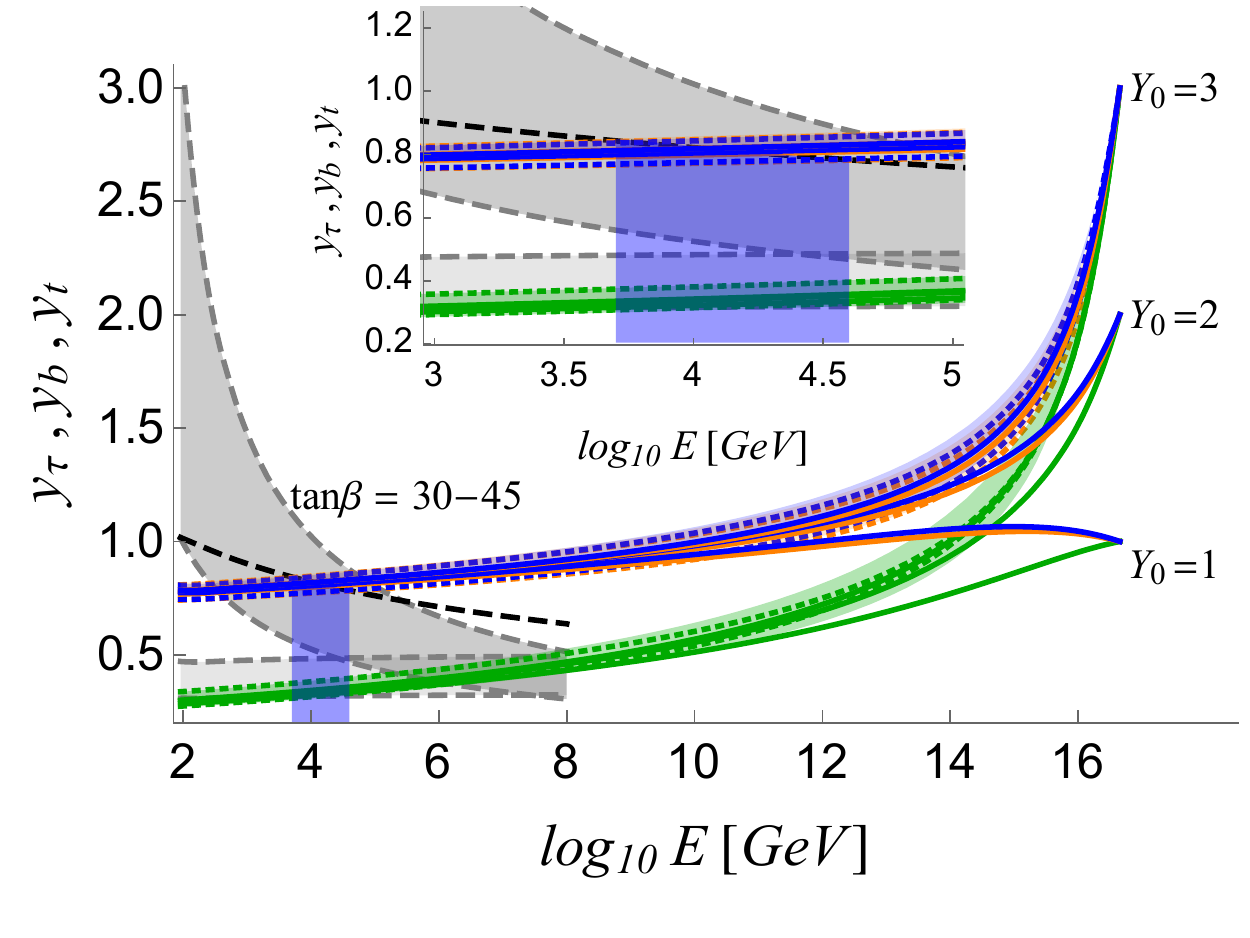}
\caption{RG evolution of $y_t$ (blue), $y_b$ (orange) and $y_\tau$ (green) in the MSSM+1VF for $\alpha_G = 0.2$ assuming three universal boundary conditions for all Yukawa couplings, $Y_0 =Y_V$. For the $Y_0=3$ the RG evolution is also shown for varying $Y_V$ between 4 and 2 (shaded ranges) and varying $\alpha_G$ in the $\pm 30\%$ range around $\alpha_G = 0.2$ (dashed lines).  No thresholds from superpartners or vectorlike matter are assumed. The  dashed lines and shaded regions at low energies show the evolution of $y_t$ (black), $y_b$ (upper gray) and $y_\tau$ (lower gray)  obtained from the measured fermion masses for $\tan\beta$ between 30 and 45  assuming that all superpartners and Higgs bosons except the SM-like one are at the corresponding RG scale.  The inset zooms in on the region at low energies.
 The blue highlight shows the range of $M$  required by $y_t$ for  the variations of  $\alpha_G $ and $Y_V$.}
\label{fig:yukawas_run}
\end{figure}

  The evolution of top, bottom and tau Yukawa couplings  in the MSSM+1VF are shown in Fig.~\ref{fig:yukawas_run} for $\alpha_G = 0.2$ and three universal boundary conditions for all Yukawa couplings, $Y_0 =Y_V$. We see that  the IR fixed point is very effective for all three Yukawa couplings since their EW scale values are barely distinguishable even in the zoomed-in plot.  In addition, for the $Y_0=3$ case, we show how little the predicted values change for order one changes in the other GUT scale parameters: the changes in the RG evolution resulting from varying $Y_V$ between 4 and 2 are indicated by shaded regions and from varying $\alpha_G$ in the $\pm 30\%$ range around $\alpha_G = 0.2$ by dashed lines.
  
\subsection{IR fixed point predictions for top, bottom and tau Yukawa couplings}

The top and bottom Yukawa couplings run to fixed ratios with respect to gauge couplings and thus their  values  at low energies are almost entirely given by the values of gauge couplings. The approximate formula for the  top Yukawa coupling can be obtained from 
\begin{equation}
 \frac{d y_{t}^2/\alpha}{dt}  = 0, \label{eq:FP}
\end{equation}
where 
\begin{equation}
\alpha \equiv \alpha_3 + \frac{9}{16} \, \alpha_2 + \frac{13}{80} \, \alpha_1
\end{equation}
 is the combination of gauge couplings, $\alpha_i = g_i^2/4\pi$,  appearing in the RG equation for $y_t$, Eq.~(\ref{eqn:RGtop}).
If we assume that all Yukawa couplings have the same boundary condition, then the only difference in the RG evolution of up-type (coupling to $H_u$) and down-type (coupling to $H_d$) couplings of quarks is due to hypercharge, see Eqs.~(\ref{eqn:RGtop}-\ref{eqn:RGbottom}), and the contribution of $y_\tau$ in the $T_{H_d}$ affecting down-type Yukawa couplings. Both of these effects are small, resulting only in tiny differences at low energies. This is the reason why the EW scale values of top and bottom Yukawa couplings in Fig.~\ref{fig:yukawas_run} are almost identical.\footnote{Similarly, since vectorlike quark Yukawa couplings have almost identical RG equations to the top and bottom Yukawa couplings up to hypercharge contributions,  assuming the same boundary conditions their RG evolution will be almost identical. Thus, we do not include them in Fig.~\ref{fig:yukawas_run}. For the same reason, the RG evolution of vectorlike lepton Yukawa couplings is almost identical to that of the tau Yukawa coupling. The IR fixed point discussion for top, bottom, and tau Yukawa couplings below equally applies to the corresponding vectorlike Yukawa couplings.} Neglecting lepton Yukawa couplings and differences from hypercharge, the IR fixed point value for the top Yukawa coupling, obtained from Eq.~(\ref{eq:FP}) and the 1-loop RG equation for top Yukawa coupling Eq.~(\ref{eqn:RGtop}), is given by: 
\begin{equation}
 \frac{s_q}{4\pi}\, y_{t,\, IR}^2  = \frac{16}{3} \alpha + \frac{2\pi}{\alpha} \frac{d \alpha}{dt} \, ,
  \label{eq:top_IR_exact}
\end{equation}
where $s_q$ is the number of $y^* y$ factors of large up-type quark Yukawa couplings. In our case $s_q = 13$. Inserting the 1-loop RG equations for gauge couplings, $d \alpha_i/dt = (b_i/2\pi) \, \alpha_i^2$, with the beta function coefficients  $b_i = (53/5,5,1)$ corresponding to the MSSM+1VF, we find
\begin{equation}
 \frac{y_{t,\, IR}^2}{4\pi}  = \frac{16}{39} \alpha + \frac{1}{\alpha} \left( \frac{1}{13} \alpha_3^2  + \frac{45}{208} \alpha_2^2 + \frac{53}{400} \alpha_1^2  \right)  .
  \label{eq:top_IR_1loop}
\end{equation}
For $\alpha_G > 0.2$ this approximation differs  from the precise numerical value by about $2\%$. Furthermore, the formula can be improved by including 2-loop terms in $d \alpha/dt$. Including just the dominant 2-loop term, proportional to $\alpha_3^3$, from the RG equation of $\alpha_3$, results in the extra term, $178 \alpha_3^3/(39\alpha)$,  on the right hand side of Eq.~(\ref{eq:top_IR_1loop}). Such an approximation agrees with the precise numerical value within  $0.5\%$. Including all 2-loop gauge terms in $d \alpha/dt$ leads to a formula that agrees with the numerical result within $0.1\%$.

\begin{figure}[t]
\centering
\includegraphics[width = 3in]{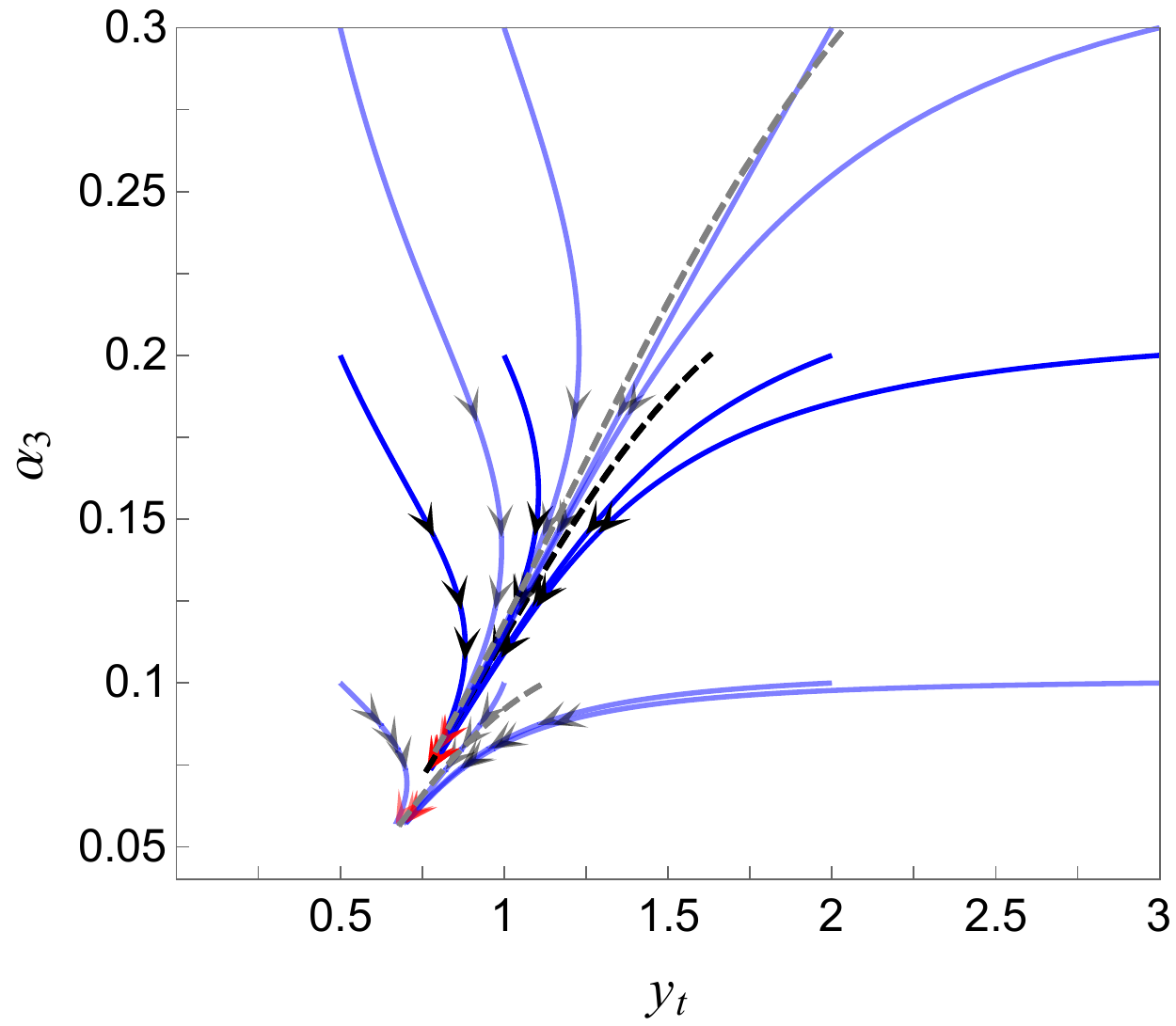}
\includegraphics[width = 3in]{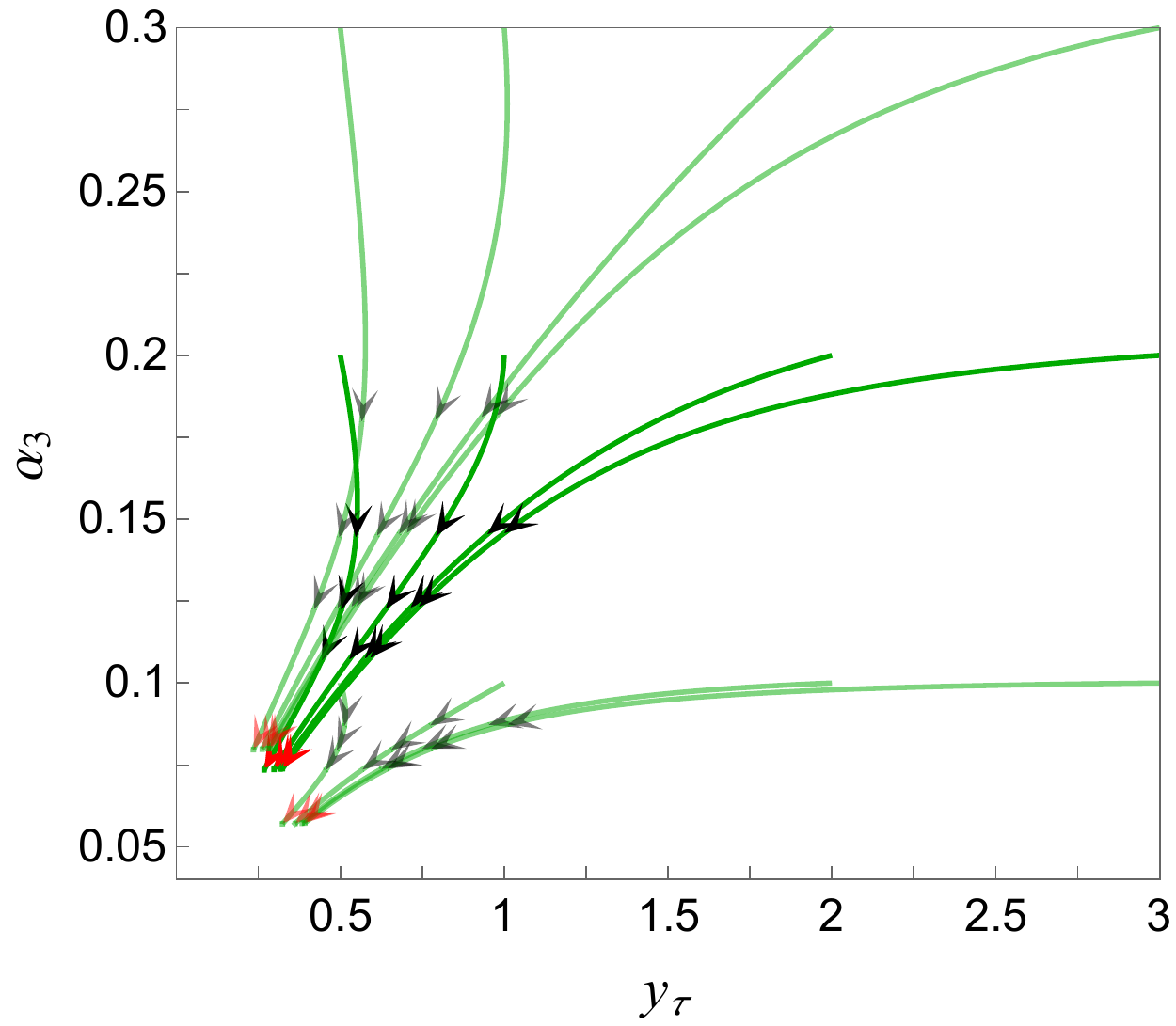}\\
\hspace{1cm}(a)\hspace{7cm} (b)
\caption{The RG flow of the top Yukawa coupling (left) and tau Yukawa coupling (right) in the $\alpha_3 - y_{t,\tau}$ planes for boundary conditions at $M_G = 3\times 10^{16}$ GeV: $\alpha_G = 0.1, \, 0.2, \,0.3$ and $Y_0 = 0.5,\, 1,\, 2, \, 3$,  assuming $Y_V = Y_0$.  The three black arrows on each line indicate values of $\alpha_3$ and $y_{t,\tau}$ at $10^{14}$ GeV, \, $10^{12}$ GeV and $10^{10}$ GeV. The last (red) arrow on each line indicates the values at the $M_Z$ scale where the RG evolution ends. Dashed lines in (a) correspond to the IR fixed point relation between  the top Yukawa coupling and gauge couplings given in Eq.~(\ref{eq:top_IR_exact}) including  2-loop gauge terms in  $d \alpha/dt$.}
\label{fig:fixed_point}
\end{figure}

These findings  indicate that 1-loop RG equations for top or bottom Yukawa couplings would be sufficient for precise predictions far below the GUT scale. However, in order to obtain the precise value of $\alpha_3$ (and thus the IR fixed point ratio for Yukawa couplings) the 2-loop terms in the RG equations of gauge couplings are needed.  The fast approach of the top Yukawa coupling to the IR fixed point from a large range of boundary conditions for $\alpha_G$ and common Yukawa coupling is  visualized in Fig.~\ref{fig:fixed_point} (left). In just about six orders of magnitude of RG running, the top Yukawa coupling is very close to the IR fixed point (dashed lines) and the IR fixed point is reached before the EW scale for any $\alpha_G > 0.1$ and $Y_0 > 0.5$.  The plotted IR fixed point relation between  the top Yukawa coupling and gauge couplings includes  2-loop gauge terms in  $d \alpha/dt$.  Since the IR fixed point value is effectively shared by large Yukawa couplings of a given type, the Eq.~(\ref{eq:top_IR_exact}) remains a very good approximation as long as $Y_0$ is comparable to $Y_V$.

It is instructive to compare different definitions of the IR fixed point of the top Yukawa coupling. The original definition, referred to as the IR stable fixed point or Pendleton-Ross fixed point~\cite{Pendleton:1980as}, corresponds to Eq.~(\ref{eq:FP}) with $\alpha$ replaced by $\alpha_3$ and using 1-loop beta function for $\alpha_3$. In our model, it would be Eq.~(\ref{eq:top_IR_1loop}) with $\alpha_{1,2}$ set to zero. However, it was realized that in practice such a value is not reached by the top Yukawa coupling in the SM (or MSSM) starting with a large boundary condition, because of a slow approach. Instead, a quasi fixed point was introduced as a value that is reached starting from large boundary conditions at the GUT scale~\cite{Hill:1980sq} (for a discussion, see also Ref.~\cite{Lanzagorta:1995gp}). Solving the 1-loop RG equations for the top Yukawa and gauge couplings we find
\begin{equation}
y_t^2(Q) = \frac{y_t^2(M_G)E(Q)}{1+\frac{s_q y_t^2(M_G)}{8\pi^2}\int_Q^{M_G} E(Q^\prime)dQ^\prime},
\end{equation}
where $E(Q)\equiv(1+\beta_3\ln(M_G/Q))^{16/3b_3}(1+\beta_2\ln(M_G/Q))^{3/b_2}(1+\beta_1\ln(M_G/Q))^{13/15b_1}$ with $\beta_i \equiv \alpha_G b_i/2\pi$.  Taking the limit of $y_t(M_G) \to \infty$ we get the formula for the quasi fixed point
\begin{equation}
y_t^2(Q) = \frac{8\pi^2E(Q)}{s_q\int_Q^{M_G}E(Q^\prime)dQ^\prime}.
 \label{eq:top_IR_Hill}
\end{equation}
As mentioned, in the SM or the MSSM the difference between Eq.~(\ref{eq:top_IR_Hill}) and Eq.~(\ref{eq:FP}) is significant  because of the slow approach to the IR fixed point and Eq.~(\ref{eq:top_IR_Hill}) is an excellent approximation of the IR fixed point value of the top Yukawa coupling. In the MSSM+1VF, the IR fixed point is approached very rapidly and the approximations based on Eq.~(\ref{eq:FP}) using 1-loop RG equations for gauge couplings and Eq.~(\ref{eq:top_IR_Hill}) agree at $0.1\%$ level for $\alpha_G > 0.2$. However, Eq.~(\ref{eq:top_IR_Hill}) is not a good approximation to the precise numerical value since 2-loop effects are sizable in the MSSM+1VF. Since  2-loop or higher order terms can be  incorporated in Eq.~(\ref{eq:FP}), this remains an excellent approximation of the IR fixed point.

As already mentioned, the evolution of the bottom Yukawa coupling is almost identical to the top Yukawa. Numerically, far away from the GUT scale, $y_b$ is typically about $0.5\%$ smaller than $y_t$. This remains to be the case also if $Y_0 \neq Y_V$.  However, the tau Yukawa coupling does not run to the IR fixed point characterized by a fixed ratio with respect to gauge couplings, but rather to the trivial IR fixed point. 

From Eq.~(\ref{eqn:RGtau}) and Fig.~\ref{fig:yukawas_run} we see that the tau Yukawa coupling is driven to smaller values by large quark Yukawa couplings that couple to $H_d$. The contributions of $\alpha_2$ and $\alpha_1$  gauge couplings and lepton Yukawa couplings are much smaller, especially far below the GUT scale. Since the quark Yukawa couplings are driven to the IR fixed point set by gauge couplings, most importantly by $\alpha_3$, also the tau Yukawa coupling is almost entirely determined by values of gauge couplings far below the GUT scale. However, as characteristic for a trivial fixed point, the value of the tau Yukawa depends also on $\alpha_G$ and how far below the GUT scale it is evaluated. 
This can be seen in Fig.~\ref{fig:fixed_point} (right). The tau Yukawa coupling approaches the same value very fast (similar to the top Yukawa coupling) from a large range of boundary conditions of Yukawa couplings, but the value is slightly different  for different $\alpha_G$. 

An insight to the general behavior of $y_\tau$ can be obtained from the RG equation of $\ln  (y_\tau^2/y_t^2) $ where we neglect everything except quark Yukawa couplings and $\alpha_3$ and approximate all quark Yukawa couplings by the top Yukawa fixed point value. 
Assuming universal GUT scale boundary conditions for Yukawa couplings and using the solution of the 1-loop RG equation of $\alpha_3$, we get 
 \begin{equation}
\frac{y_{\tau}^2 (M_Z)}{y_{t}^2(M_Z)} \; \sim \;  \frac{1}{\left[1+(\alpha_G/2\pi) \ln(M_G/M_Z) \right]^{132/39}}.
\end{equation}
As anticipated, the ratio of $y_{\tau}^2$ to quark Yukawa couplings squared (or to gauge couplings) is decreasing for  larger $\alpha_G$ and  further away from the GUT scale it is evaluated.\footnote{The formula above is just a very rough approximation intended for the illustration of general behavior of $y_\tau$. It is not suitable as an approximation of the actual value of $y_\tau$ at the EW scale. The effects of other gauge couplings and lepton Yukawa couplings in the RG flow are not negligible.}

So far we have not included the threshold effects from superpartners or vectorlike matter, the same particle content is assumed all the way to the EW scale. As a result, the gauge couplings do not reproduce the measured values exactly. This is intentional since we want to infer the scale of superpartners and vectorlike matter from the IR fixed point values of Yukawa couplings. In order to do this, with dashed lines and shaded regions at low energies in Fig.~\ref{fig:yukawas_run}, we plot  the evolution of $y_t$ (black), $y_b$ (upper gray) and $y_\tau$ (lower gray)  obtained from the measured fermion masses for $\tan\beta$ between 30 and 45 assuming that all superpartners and Higgs bosons  except the SM-like one are at the corresponding RG scale. They are obtained from Eqs.~(\ref{eq:yt-corrections}) - (\ref{eq:ytau-corrections}) by identifying the $M_{SUSY}$ scale  with the RG scale.\footnote{Note that these do not represent RG evolutions of Yukawa couplings in any model (neither in the MSSM nor in the SM), but rather evolutions of the combination of model parameters that have to match Yukawa couplings in the MSSM+1VF at the correct scale of new physics in order to obtain measured fermion masses. We use  the crossing point of the RG evolution of this quantity and the corresponding Yukawa coupling in the MSSM+1VF  to infer $M$.} For the top quark, the whole region of $\tan \beta \gtrsim 10$ is essentially the same line and thus the top quark IR fixed point is the most restrictive on  the scale of superpartners and vectorlike matter. The range of $M$ required by the top quark mass for the above mentioned variations of GUT scale parameters is indicated by the vertical shaded band.

Interestingly, the multi-TeV range of $M$  suggested by the top quark mass coincides with the range already suggested by the gauge couplings~\cite{Dermisek:2017ihj} and is also compatible with the Higgs boson mass. In addition, predicted values of the bottom and tau Yukawa couplings in this energy range are within gray shaded regions indicating that the resulting bottom and tau masses will not be far from measured values for $\tan \beta \sim 40$.

\subsection{Fits to low energy observables and the scale of new physics}

 \begin{figure}[t]
\centering
\includegraphics[width = 3in]{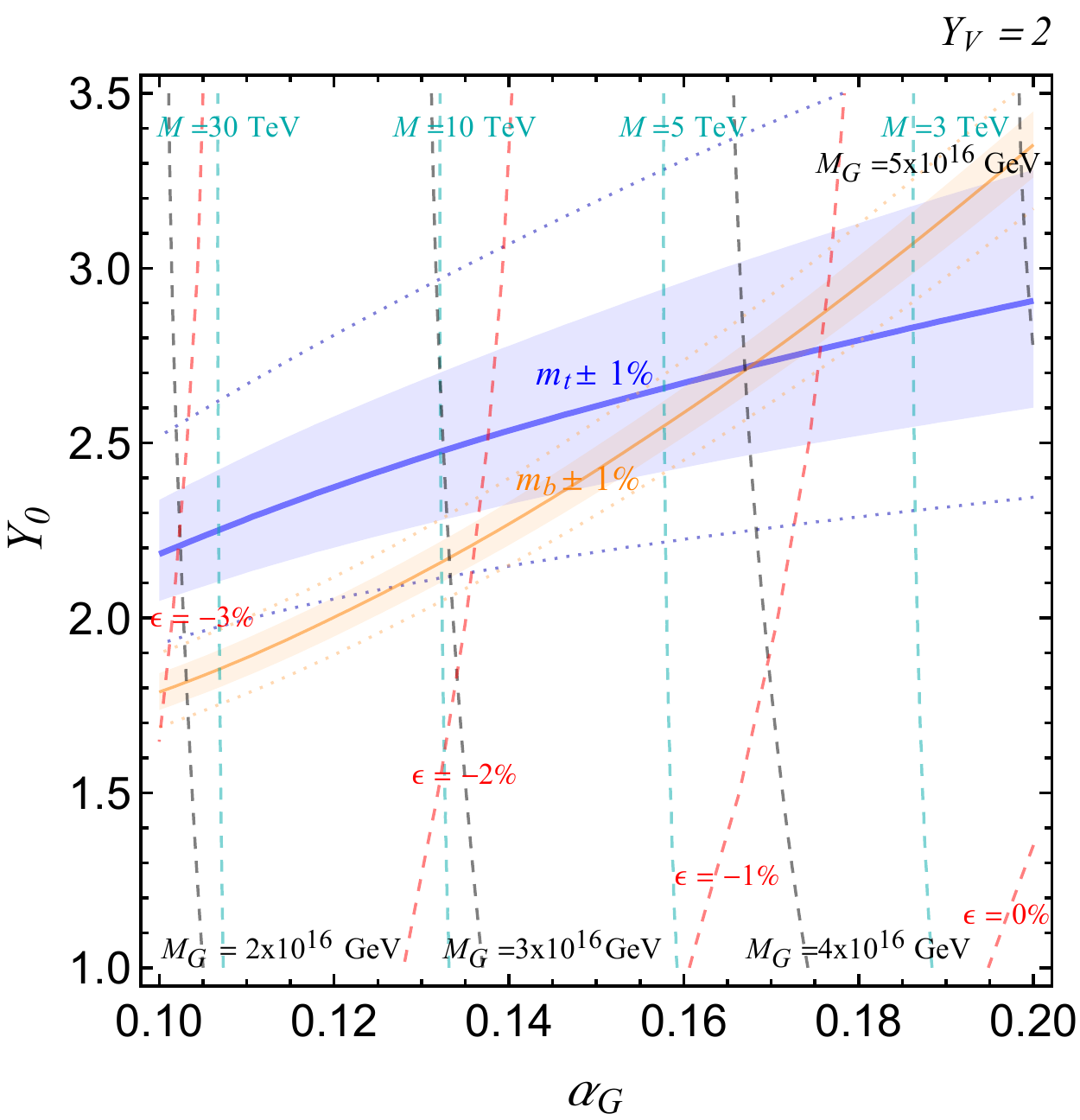}
\includegraphics[width = 3in]{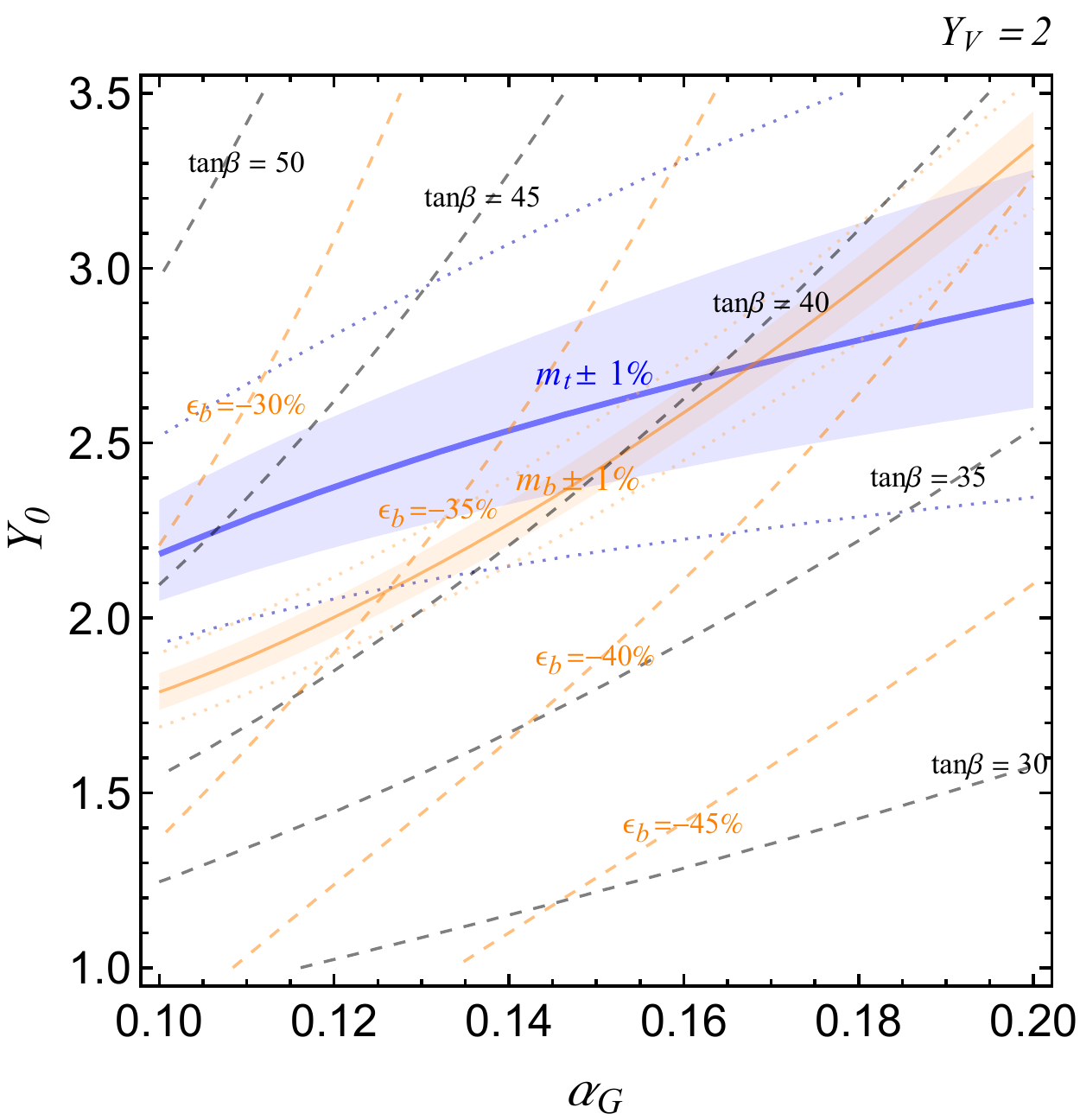}\\
\hspace{1cm}(a)\hspace{7cm} (b)
\caption{Contours of predicted $m_t$ (blue) and $m_b$ (orange) in the $Y_0 - \alpha_G$ plane for fixed $Y_V = 2$. Solid  lines correspond to the measured central values, shaded areas represent $\pm 1\%$ ranges and dotted lines correspond to $\pm 2\%$ ranges. The $m_\tau$ and all three gauge couplings are fit to the measured central values everywhere in the plane for the values of input parameters plotted with dashed lines: $\epsilon$, $M_G$ and $M$ in (a) and  $\tan \beta$ in (b). In (b) we also show contours of constant $\epsilon_b$ that would be required to obtain the measured value of $m_b$.}
\label{fig:Y0_aG}
\end{figure}

The next step is to determine the region of model parameters leading to exact Yukawa coupling unification  while still keeping a high degree of universality in model parameters $M\equiv M_{SUSY} = M_{VF}$. One of the fermion masses can  always be reproduced precisely for some value of $\tan \beta$. Since the tau mass receives only small corrections from superpartners and is  known the most precisely we choose to fix $\tan \beta$ to obtain the central value of $m_\tau$. In Fig.~\ref{fig:Y0_aG} we then plot the contours of predicted $m_t$ and $m_b$ in the $Y_0 - \alpha_G$ plane along which the measured central values, $\pm 1\%$ and $\pm 2\%$ ranges are obtained.  All three gauge couplings are fit to their central values for the values of $M_G$, $M$ and $\epsilon$ plotted with dashed lines in (a). Contours of constant $\tan \beta$ required to fit $m_\tau$ are shown in (b). Since $m_b$ is the most sensitive to SUSY spectrum, in (b) we also indicate values of $\epsilon_b$ that would be required to obtain the measured value of $m_b$ everywhere in the $Y_0 - \alpha_G$ plane.

 \begin{figure}[t]
\centering
\includegraphics[width = 3in]{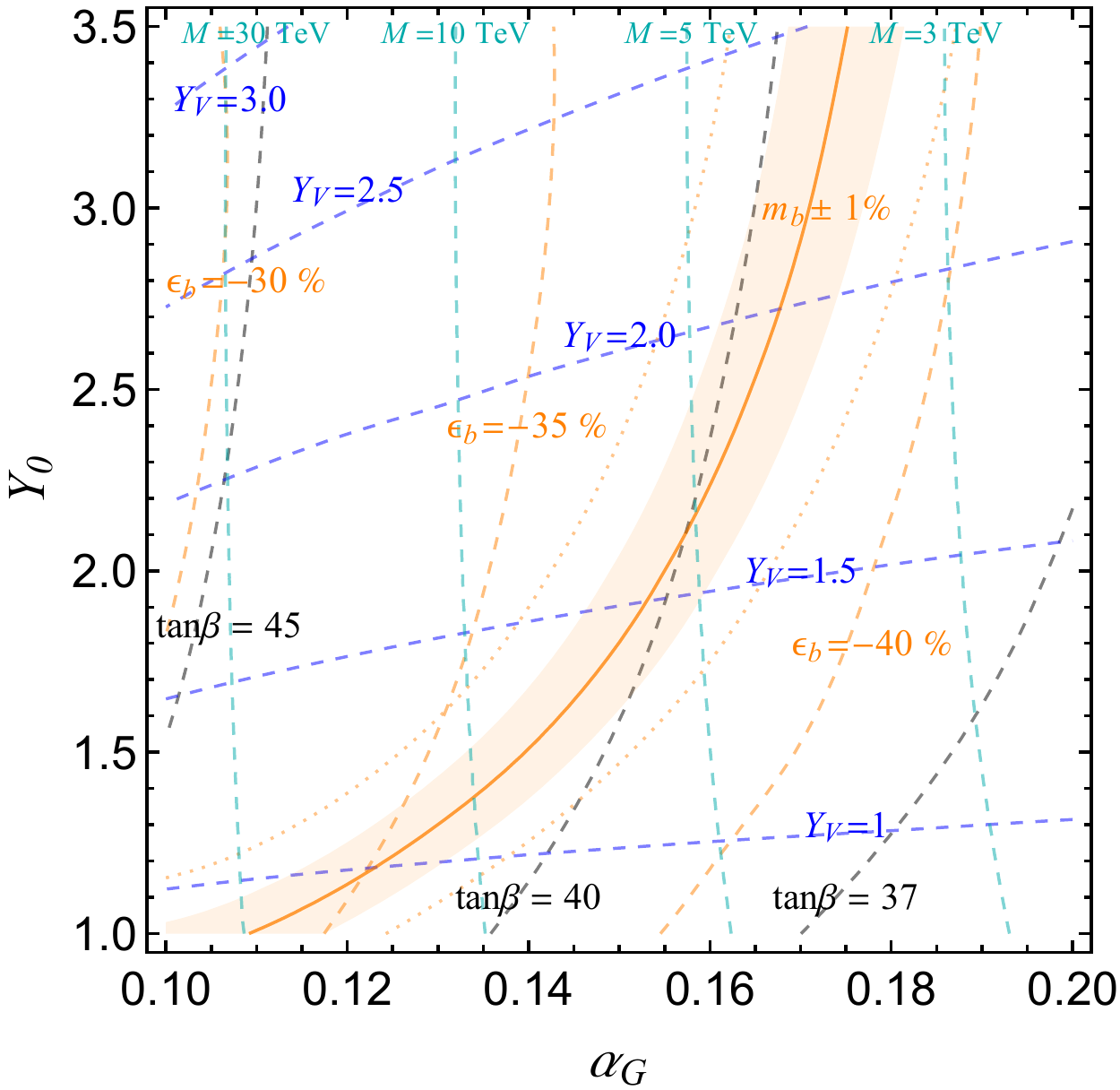}
%\includegraphics[width = 3in]{MV_MSUSY_Y0_3_8_tb_40}\\
%\hspace{1cm}(a)\hspace{7cm} (b)
\caption{Similar to Fig.~\ref{fig:Y0_aG} but with the top quark mass also fit to the measured central value everywhere in the plane for values of $Y_V$ indicated by the dashed blue lines. }
\label{fig:Y0_aG_YV_mb}
\end{figure}

From Fig.~\ref{fig:Y0_aG} we see that the third generation Yukawa couplings successfully unify for  values of model parameters  near the crossing of the solid lines corresponding to central values of top and bottom masses. Note, the current experimental uncertainty for the top quark mass is about half of the shaded region while for the bottom quark it coincides with the shaded range.  In this plot $Y_V$ is fixed to 2. Different choices of $Y_V$ would slowly move the region where top and bottom quark masses are correctly reproduced up and down in the plane. Alternatively, we can use $Y_V$ to fit the central value of the top quark mass everywhere in the plane. Contours of the predicted bottom quark mass with all other observables fit to central values are shown in Fig.~\ref{fig:Y0_aG_YV_mb}. We also show contours of the required $Y_V$ and a subset of other model parameters. Those not shown have similar values as in Fig.~\ref{fig:Y0_aG}.

Perhaps more interesting than the Yukawa coupling unification itself is the fact that the unification is possible with large boundary conditions for both gauge and Yukawa couplings. It means that the EW scale values are very insensitive to the boundary conditions due to the IR fixed point behavior discussed above. This can also  be inferred from the large size of the parameter space leading to the top quark mass in 1\% or 2\% ranges around the central value in Fig.~\ref{fig:Y0_aG}. Moreover, from the shaded range of the bottom quark mass in Fig.~\ref{fig:Y0_aG_YV_mb} we see that for Yukawa couplings larger than one, the preferred range of superpartners and vectorlike matter is  3 TeV to 30 TeV,  with larger couplings favoring smaller scales of new physics. This mass range is also compatible with the Higgs boson mass.

Another interesting feature is that the required SUSY correction to $m_b$ in the whole plotted plane  is in the range that is generically achieved with comparable values of SUSY parameters. Thus, no extreme regions of SUSY parameters are required to simultaneously obtain all three fermion masses correctly. This is important, since due to the IR fixed point nature, there are no other parameters that can affect the fermion masses significantly.

 \begin{figure}[t]
\centering
\includegraphics[width = 3in]{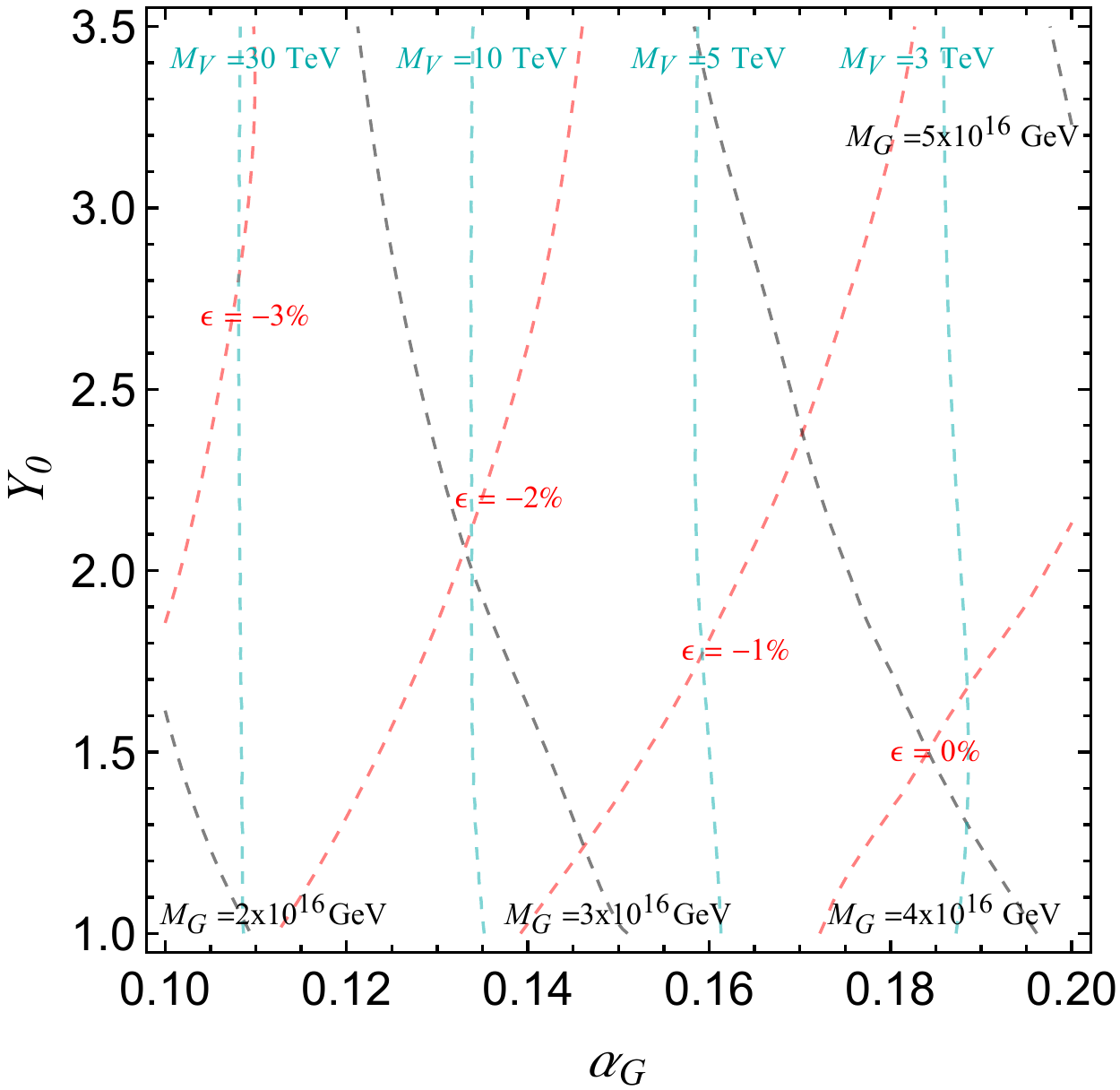}
\includegraphics[width = 3in]{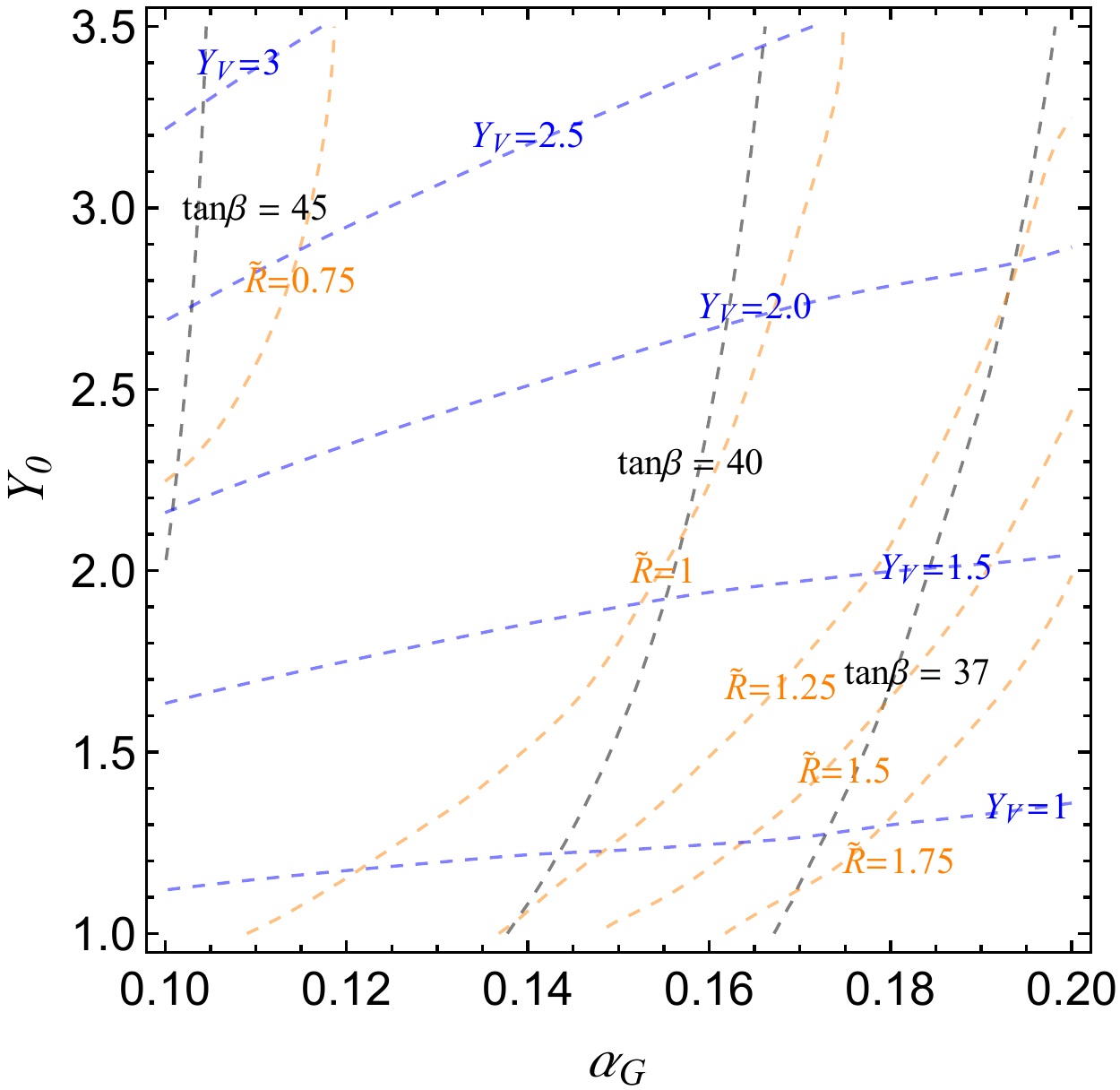}\\
\hspace{1cm}(a)\hspace{7cm} (b)
\caption{Contours of constant values of model parameters required to fit the central values of three gauge couplings and third generation fermion masses in the $Y_0 - \alpha_G$ plane. Parameters mostly related to gauge couplings are shown in (a) and those mostly related to fermion masses are shown in (b). The $\tilde R$ is the ratio of gaugino masses and  scalar masses with scalar masses set to  $M_V$.}
\label{fig:Y0_aG_R}
\end{figure}

Splitting gaugino and scalar masses  can actually be used to fit the bottom quark mass everywhere in the plane. Defining $\tilde R$ parameter as the ratio of gaugino masses and scalar masses and still, for simplicity, assuming that scalar masses are the same as vectorlike quark and lepton masses, this is illustrated in Fig.~\ref{fig:Y0_aG_R} which shows the model parameters required to fit the central values of all three gauge  couplings and three fermion masses everywhere in the $Y_0 - \alpha_G$ plane. We see that splitting gaugino and scalar masses by less than a factor of two is sufficient to get all three fermion masses at central values from the unified Yukawa coupling everywhere in the plane.   No extreme regions of SUSY parameters or large hierarchies are required.

 \begin{figure}[t]
\centering
\includegraphics[width = 3in]{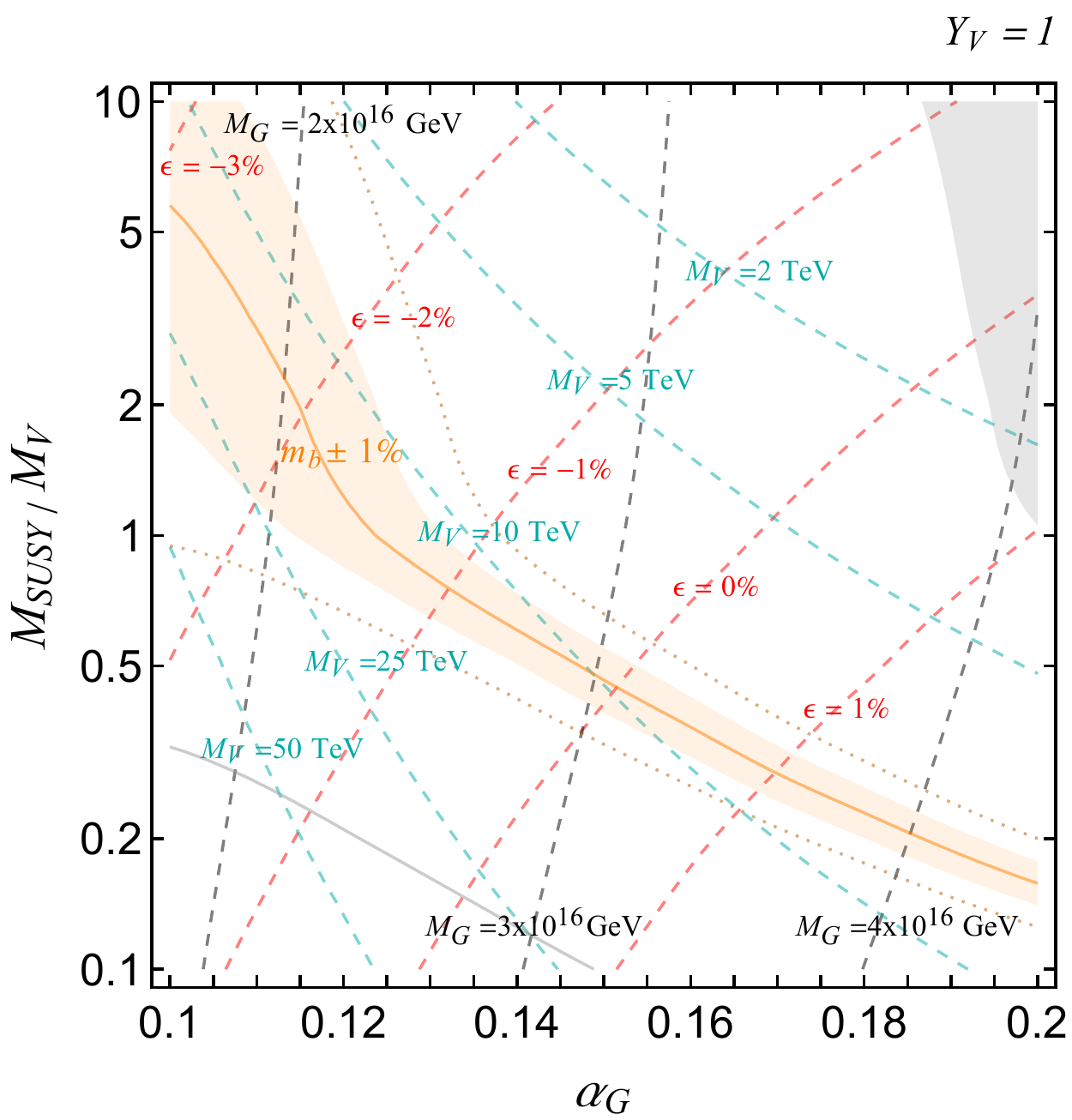}
\includegraphics[width = 3in]{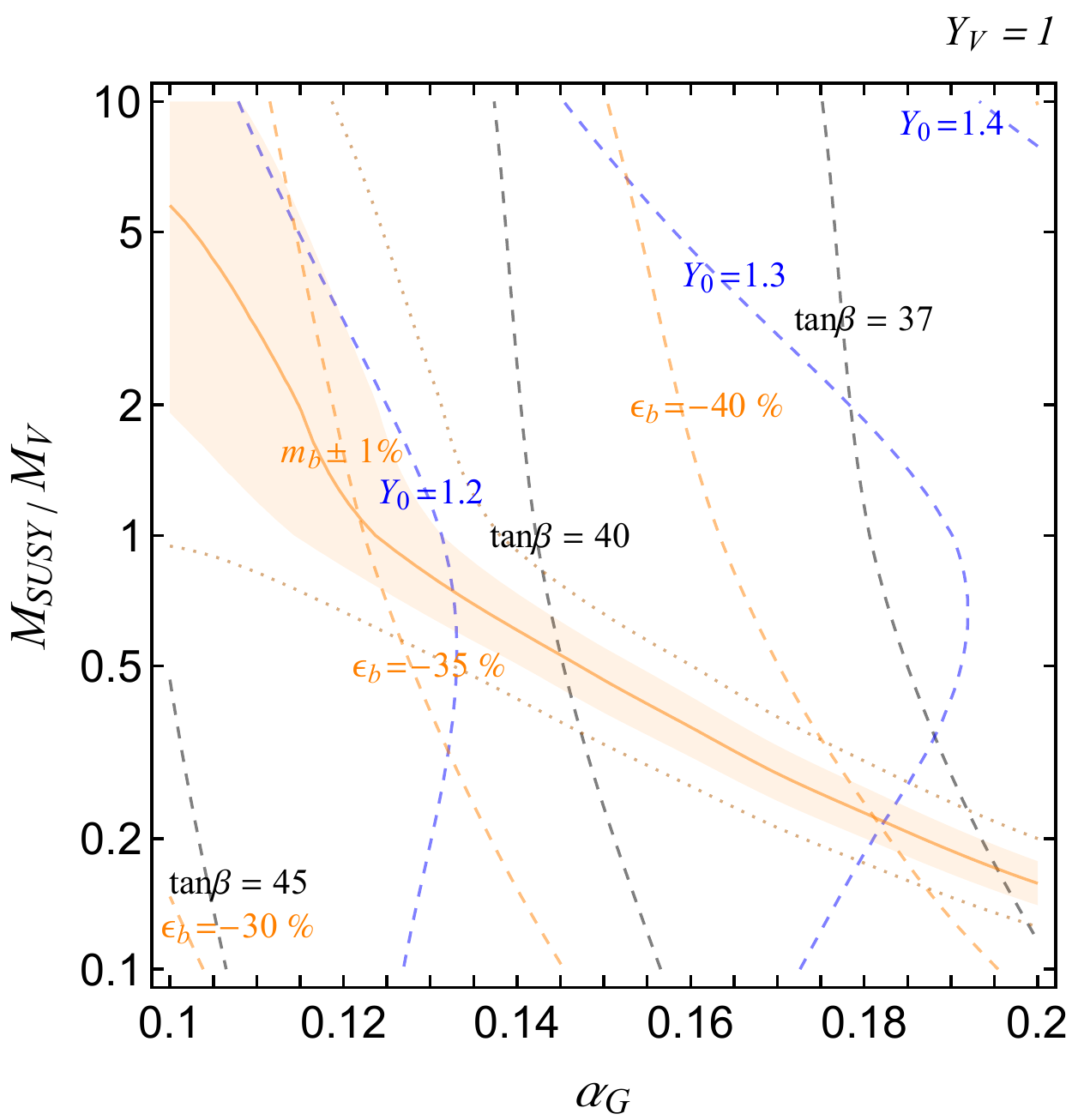}\\
\hspace{1cm}(a)\hspace{7cm} (b)
\caption{Contours of predicted $m_b$ (orange) in the $M_{SUSY}/M_V - \alpha_G$ plane for fixed $Y_V = 1$. Solid  line corresponds to the measured central value, shaded area represent $\pm 1\%$ range and dotted lines correspond to $\pm 2\%$ range. The $m_t$, $m_\tau$ and all three gauge couplings are fit to the measured central values everywhere in the plane for the values of input parameters plotted with dashed lines:  $M_G$, $M$ and $\epsilon$ in (a) and  $Y_0$ and $\tan \beta$ in (b). In (b) we also show contours of constant $\epsilon_b$ that would be required to obtain the measured value of $m_b$. The edge of the shaded gray area in (a) corresponds to the contour of measured central value of $m_b$ for $Y_V = 2.5$ (that requires $Y_0 \simeq 3.5$) and the gray solid line in the bottom left corner  corresponds to $m_b$ for $Y_V = 0.5$. }
\label{fig:RM_aG}
\end{figure}

Finally, let us explore the effects of splitting the common scale of superpartners, $M_{SUSY}$, from the common scale of vectorlike masses, $M_{VF}$. These effects are generically very mild unless the level of splitting is huge. Even an order of magnitude changes in  $M_{SUSY}$ or $M_{VF}$  have only a tiny impact on $m_t$, at most a few percent effect on $m_b$ and of order 10\% effect on $m_\tau$ when other parameters are fixed.  Instead of showing this we can use the ratio $M_{SUSY}/M_{V}$ as a free parameter and see how other parameters have to compensate for this change in order to fit fermion masses. In Fig.~\ref{fig:RM_aG} we plot the predicted $m_b$  in the $M_{SUSY}/M_{V} -\alpha_G$ plane together with contours of model parameters required  to fit the central values of top and tau masses in addition to gauge couplings.  In these plots we fix $Y_V = 1$. Larger values of $Y_V$ would shift  the $m_b$  contour slightly up and smaller values slightly down. The position of the correct bottom quark mass for different $Y_V$ can be easily  estimated by comparing with  Fig.~\ref{fig:Y0_aG_YV_mb}.
For guidance, the edge of the shaded gray area in Fig.~\ref{fig:RM_aG}(a) corresponds to the contour of the measured central value of $m_b$ for $Y_V = 2.5$ (and requires $Y_0 \simeq 3.5$). Inside the gray shaded region the $Y_0$ required to fit the central value of the bottom quark mass grows rapidly. The gray solid line in the bottom left corner corresponds to  the measured central value of  $m_b$ for $Y_V = 0.5$. We show these additional contours only in Fig.~\ref{fig:RM_aG}(a) since the model parameters displayed there are driven mostly by gauge couplings and depend negligibly on $Y_V$.\footnote{Note, however, that large Yukawa couplings result in an improvement of gauge coupling unification. The required GUT scale threshold correction is significantly smaller than without extra Yukawa couplings \cite{Dermisek:2017ihj}. The improvement is mostly the result of contributions of Yukawa couplings to $\alpha_1$ and $\alpha_2$ over the whole range of RG evolution and depends very little on the boundary conditions as long as they are large.} The parameters shown in Fig.~\ref{fig:RM_aG}(b) are mostly related to fermion masses and would be affected by changing $Y_V$.

We see that predictions for fermion masses indeed do not depend much on how the scale of  superpartners is split from vectorlike masses. Exact Yukawa coupling unification can be achieved in large ranges of  $M_{SUSY}$ and $M_{V}$  for slightly different values of unified gauge and Yukawa couplings.  Both $M_{SUSY}$ and $M_{V}$ are preferred in a multi-TeV range that starts from about 2 TeV for large values of $Y_0$, $Y_V$ and $\alpha_G$. Furthermore, the required $M_{SUSY}$ and $M_{V}$ are somewhat anti-correlated: smaller $M_{SUSY}$ prefers larger $M_{V}$ and vice versa. These finding are mostly driven by fitting the measure values of gauge couplings~\cite{Dermisek:2017ihj}, with fermion masses further constraining the preferred range.   Lowering Yukawa couplings to 1, the range of  $M_V$ extends  to about 8 TeV for $\alpha_G > 0.2$ and up to about 45 TeV for $\alpha_G = 0.1$ while $M_{SUSY}$ can be an order of magnitude smaller or larger depending on $\alpha_G$. For any $\alpha_G>0.2$ either superpartners or vectorlike matter is expected within 3 TeV. Further decreasing the Yukawa  couplings (or $\alpha_G$) requires larger scales of new physics. However, in this limit, the understanding of the third generation fermion masses as IR fixed points gradually fades away. We should keep in mind, however, that these results assume the typical SUSY corrections resulting from comparable SUSY spectrum and different assumptions about soft SUSY breaking terms or the $\mu$-term could shift the preferred range of model parameters as indicated (by $\epsilon_b$) in Fig.~\ref{fig:RM_aG}(b) and previous figures.

 \section{Conclusions}
 \label{sec:conclusions}

We have found  that in the MSSM extended by a complete vectorlike family, precise top, bottom and tau Yukawa coupling unification can be achieved with a large unified coupling, implying that all three fermion masses can be simultaneously close to their IR fixed points. All three Yukawa couplings approach IR fixed points rapidly from a large range of boundary conditions both above and below their IR fixed point values. Furthermore, the unification is possible assuming SUSY threshold corrections which are typical for comparable superpartner masses and thus no hierarchies or specific relations among SUSY parameters are required. 

The simplest scenario assumes a common scale of new physics (superpartner masses and masses of vectorlike fermions). This scale, together with the GUT scale and $\tan \beta$ are the most important parameters determining the EW scale values of the top, bottom and tau Yukawa couplings while others affect the EW scale values very little, as a result of the IR fixed point behavior,  and are only needed for precisely reproducing the measure values.  For 
 unified Yukawa couplings of order one or larger, the preferred scale of superpartners and vectorlike matter is in the 3 TeV to 30 TeV range, with larger couplings favoring smaller scales of new physics. The required scale of new physics and the GUT scale are to a large extent  driven by fitting the measure values of gauge couplings~\cite{Dermisek:2017ihj} with fermion masses further constraining the preferred range. Due to the IR fixed point behavior it is highly non-trivial that  Yukawa couplings point to a similar scale of new physics as gauge couplings. Furthermore, the multi-TeV scale for superpartners  is independently suggested by  the Higgs boson mass.

Abandoning the simple assumption of a common scale of new physics, the results do not differ much as long as superpartner masses and vectorlike masses remain comparable.
Both $M_{SUSY}$ and $M_{V}$ are preferred in a multi-TeV range that starts from about 2 TeV for large values of unified Yukawa couplings  and $\alpha_G$.  Lowering Yukawa couplings to 1, the range of  $M_V$ extends  to about 8 TeV for $\alpha_G > 0.2$ and up to about 45 TeV for $\alpha_G = 0.1$ while $M_{SUSY}$ can be an order of magnitude smaller or larger depending on $\alpha_G$. For any $\alpha_G>0.2$ either superpartners or vectorlike matter is expected within  3 TeV. Further decreasing the Yukawa  couplings (or $\alpha_G$) requires larger scales of new physics. However, in this limit, the understanding of the third generation fermion masses as IR fixed points gradually fades away.  

The above motivation for the scale of superpartners and vectorlike matter is based completely on the measured values of the third generation fermion masses together with gauge couplings  and does not take into account any  biases related to naturalness of EW symmetry breaking. It coincides with the only hint  for the scale of superpartners we have so far (the Higgs boson mass). Not assuming any specific SUSY breaking/mediation model, many scenarios with basic features similar to those we considered are sufficiently complex that the needed hierarchy between the EW scale and the scale of new physics does not require model parameters to be selected with any special care~\cite{Dermisek:2016zvl, Dermisek:2017xmd}.

Although the preferred scale  of superpartners  and vectorlike matter is in a multi-TeV range, any of the new particles can be within the reach of the LHC since the prediction for a Yukawa or gauge coupling depends on  a weighted geometric mean of masses of particles contributing in its RG evolution. Based on the RG evolution, vectorlike leptons and both MSSM and vectorlike sleptons are expected at the bottom of the spectrum. Similarly, extra Higgs bosons resulting from the two Higgs doublets may be light. Thus, in addition to the usual searches for either vectorlike matter or heavy Higgs bosons, combined signatures of both sectors are of particular importance since the sensitivity for some of those extends to several TeV~\cite{Dermisek:2015hue, Dermisek:2016via}.

 The model we consider is certainly more complex than the MSSM. Nevertheless, it offers a simple understanding of gauge and the third generation Yukawa couplings  in terms of scales of new physics.

%%%%%%%%%%%%%%%%%%%%%%%%%%%%%%%%%%%%%%%%%%%%%%%%%%%%%%%%
%\acknowledgements
%%%%%%%%%%%%%%%%%%%%%%%%%%%%%%%%%%%%%%%%%%%%%%%%%%%%%%%%

\vspace{0.2cm}
\noindent
{\bf Acknowledgments:} This work was supported in part by the U.S. Department of Energy under grant number {DE}-SC0010120.

%%%%%%%%%%%%%%%%%%%%%%%%%%%%%%%%%%%%%%%%%%%%%%%%%%%%%%%%

%%%%%%%%%%%%%%%%%%%%%%%%%%%%%%%%%%%%%%%%%%%%%%%%%%%%%%%%%

%%%%%%%%%%%%%%%%%%%%%%%%%%%%%%%%%%%%%%%%%%%%%%%%%%%%%%%%%
\end{document}